\documentclass[sigconf]{acmart}

\AtBeginDocument{%
  }

\copyrightyear{2026}
\acmYear{2026}
\setcopyright{cc}
\setcctype{by}
\acmConference[KDD '26]{Proceedings of the 32nd ACM SIGKDD Conference on Knowledge Discovery and Data Mining V.2}{August 09--13, 2026}{Jeju Island, Republic of Korea}
\acmBooktitle{Proceedings of the 32nd ACM SIGKDD Conference on Knowledge Discovery and Data Mining V.2 (KDD '26), August 09--13, 2026, Jeju Island, Republic of Korea}
\acmDOI{10.1145/3770855.3817526}
\acmISBN{979-8-4007-2259-2/2026/08}

\usepackage{microtype}
\usepackage{graphicx}
\usepackage{subfigure}
\usepackage{booktabs} 

\usepackage{hyperref}
\usepackage{xspace}
\usepackage{float}
    
\usepackage{amsmath}
\usepackage{mathtools}
\usepackage{multicol}
\usepackage{multirow}
\usepackage{xcolor}

\usepackage{pgf,amsmath,tikz}
\usetikzlibrary{shapes.misc, positioning}
\usetikzlibrary{shapes.geometric}
\usetikzlibrary{shadows}
\usetikzlibrary{arrows}

\usepackage[capitalize,noabbrev]{cleveref}

\usepackage[textsize=tiny]{todonotes}
\usepackage{tcolorbox}

\newcommand{\cmd}[1]{{\ttfamily #1}}

\newcommand{\java}{{\cmd{Java}}}

\newcommand{\github}{{\cmd{Github}}}

\newcommand{\claude}{{\cmd{Claude-4.5-Sonnet}}}

\newcommand{\javan}[1]{{\cmd{Java $#1$}}}

\newcommand{\mvnverify}{{\cmd{mvn clean verify}}}
\newcommand{\mvncompile}{{\cmd{mvn clean compile}}}
\newcommand{\mvn}{{\cmd{mvn}}}

\newcommand{\maven}{{\cmd{maven}}}

\newcommand{\etamin}{\text{minimal}}
\newcommand{\etamax}{\text{maximal}}
\newcommand{\compile}{\text{compile}}

\newcommand{\seedchange}{$\Delta_{\scriptsize \text{seed}}$}
\newcommand{\footnoteurl}[1]{\footnote{\;\url{#1}}}
\newcommand{\underlinejv}[1]{#1}

\newcommand{\textlite}{selected}
\newcommand{\lite}{\cmd{\textlite}}
\newcommand{\migrationbench}{{MigrationBench}}
\newcommand{\migrationfull}{\cmd{migration-bench-full}}

\newcommand{\migrationlite}{\cmd{migration-bench-\textlite}}

\newcommand{\repo}{repository}
\newcommand{\repos}{repositories}
\newcommand{\filter}[1]{$\mathbf{F_{#1}}$}

\newcommand{\white}[1]{{\color{white}#1}}

\newcommand{\hidezero}{\white{.0}}

\newcommand{\dblue}[1]{#1}

\newcommand{\huggingFaceCollectionUrl}{https://huggingface.co/collections/AmazonScience/migrationbench}
\newcommand{\githubUrl}{https://github.com/amazon-science/MigrationBench}

\newcommand{\mineta}{71.67}
\newcommand{\maxeta}{53.33}

\usepackage{array}
\usetikzlibrary{shapes.multipart}

\usepackage[utf8]{inputenc} 
\usepackage[T1]{fontenc}    
\usepackage{hyperref}       
\usepackage{url}            
\usepackage{booktabs}       
\usepackage{amsfonts}       
\usepackage{nicefrac}       
\usepackage{microtype}      
\usepackage{xcolor}         
\usepackage{verbatim}
\usepackage{listings}
\usepackage{xcolor}
\usepackage{threeparttable} 
\usepackage{threeparttablex}

\newtoggle{arxivLatex}
\newtoggle{arxivLatexRef}
\toggletrue{arxivLatexRef}
\togglefalse{arxivLatexRef}

\toggletrue{arxivLatex}

\iftoggle{arxivLatex}{
}{
\usepackage{minted}
}

\usepackage{listings}
\usepackage[compatibility=false]{caption}
\usepackage{xcolor}

\definecolor{Box1Color}{RGB}{227, 236, 246}
\definecolor{Box2Color}{RGB}{248, 220, 225}
\definecolor{Box3Color}{RGB}{255, 238, 224}
\definecolor{Box4Color}{RGB}{235, 255, 240} 
\definecolor{Box5Color}{RGB}{249, 238, 236}

\definecolor{Box4Color30}{RGB}{244, 255, 247}
\definecolor{Box4Color50}{RGB}{245, 255, 247}
\definecolor{Box4Color70}{RGB}{249, 255, 250}
\definecolor{Box4Color90}{RGB}{253, 255, 254}

\title{\migrationbench: Repository-Level Code Migration Benchmark from \cmd{Java} $8$}

\author{Linbo Liu}
\authornote{Both authors contributed equally to this research.}
\affiliation{%
  \institution{AWS AI}
  \city{Santa Clara, CA}
  \country{USA}}
\email{linbol@amazon.com}

\author{Xinle Liu}
\authornotemark[1]
\affiliation{%
  \institution{AWS AI}
  \city{Santa Clara, CA}
  \country{USA}}
\email{sliuxl@amazon.com}

\author{Qiang Zhou}
\affiliation{%
  \institution{AWS AI}
  \city{Sunnyvale, CA}
  \country{USA}}
\email{zhouqia@amazon.com}

\author{Lin Chen}
\affiliation{%
  \institution{AWS AI}
  \city{Santa Clara, CA}
  \country{USA}}
\email{linchenk@amazon.com}

\author{Yihan Liu}
\affiliation{%
  \institution{AWS AI}
  \city{Santa Clara, CA}
  \country{USA}}
\email{yyihanl@amazon.com}

\author{Hoan Nguyen}
\affiliation{%
  \institution{AWS AI}
  \city{Seattle, WA}
  \country{USA}}
\email{hoanamzn@amazon.com}

\author{Behrooz Omidvar-Tehrani}
\affiliation{%
  \institution{AWS AI}
  \city{Santa Clara, CA}
  \country{USA}}
\email{omidvart@amazon.com}

\author{Xi Shen}
\affiliation{%
  \institution{AWS AI}
  \city{Santa Clara, CA}
  \country{USA}}
\email{xis@amazon.com}

\author{Jun Huan}
\affiliation{%
  \institution{AWS AI}
  \city{Santa Clara, CA}
  \country{USA}}
\email{lukehuan@amazon.com}

\author{Omer Tripp}
\affiliation{%
  \institution{AWS AI}
  \city{Sunnyvale, CA}
  \country{USA}}
\email{omertrip@amazon.com}

\author{Anoop Deoras}
\affiliation{%
  \institution{AWS AI}
  \city{Santa Clara, CA}
  \country{USA}}
\email{adeoras@amazon.com}

\begin{document}

\renewcommand{\shortauthors}{Liu et al.}

\begin{abstract}


With the rapid advancement of powerful large language models (LLMs) in recent years, a wide range of software engineering tasks can now be addressed using LLMs, significantly enhancing productivity and scalability.
Numerous benchmark datasets have been developed to evaluate the coding capabilities of these models, while they primarily focus on code generation
and issue-resolution tasks.
In contrast, 
we introduce a new coding benchmark \migrationbench\ with a distinct focus: code migration.
\migrationbench\ aims to serve as a comprehensive benchmark for migration from \javan{8} to the latest long-term support (LTS) versions (\javan{17}, $21$), 
including a
\href{https://huggingface.co/datasets/AmazonScience/migration-bench-java-full}{\cmd{full}}
dataset and its subset
\href{https://huggingface.co/datasets/AmazonScience/migration-bench-java-selected}{\lite} with 
$5,102$ and $300$ \repos\ respectively.
\lite\ is a representative subset curated for complexity and difficulty,
offering a versatile resource to support research in the field of code migration. 
Additionally,
we provide a comprehensive evaluation framework to facilitate rigorous and standardized assessment of LLMs on this challenging task.
We further propose an agentic framework and demonstrate that LLMs can effectively tackle repository-level code migration to \javan{17}. For the \lite\ subset with \claude, our agentic framework achieves $\mineta\%$  and $\dblue{\maxeta}\%$ success rate (pass@1) for minimal and maximal migration respectively. 
The dataset and evaluation source code are available at: 
\href{\huggingFaceCollectionUrl}{https://huggingface.co/collections/AmazonScience/migrationbench}
and \url{\githubUrl} respectively.

\end{abstract}





\maketitle


\section{Introduction}
\label{sec:intro}

The rapid development of large language models (LLMs) with billions of parameters has led to an unprecedented growth in their size and reasoning capabilities~\citep{kaplan2020scaling, wei2022emergent, ganguli2022predictability}. The software engineering (SWE) industry involves various coding tasks with relatively objective goals~\cite{liu2024large}, requiring strong domain knowledge and reasoning abilities. Consequently, the research community has been actively and effectively leveraging LLMs in this domain~\cite{schafer2023empirical, chen2023teaching}. Initially, LLMs were employed for standalone and interview-style code generation tasks~\cite{mbpp, hendrycks2021measuring, humaneval}.
For instance,
GitHub Copilot\footnoteurl{https://github.com/features/copilot}
assists in completing functions within Integrated Development Environments (IDEs). Subsequently, their applications expanded to more complex use cases,
like SWE-Bench~\cite{Jimenez2024}, which generates pull requests to address open issues for GitHub repositories.


As LLMs demonstrated remarkable success in these coding tasks~\cite{xia2024agentless, huang2023agentcoder, shinn2024reflexion, vijay2025enhancing}, attention has shifted to even more complex challenges in software engineering, such as code migration~\cite{nikolov2025google, bairi2024codeplan}. Unlike code generation or issue resolution, which often focuses on isolated functions or files, code migration tasks require a holistic approach involving the entire repository. Migration entails navigating a multi-step process to address numerous interconnected issues across files, making it a far more intricate task~\cite{kontogiannis2010code}. This complexity highlights the need to evaluate LLMs on such advanced use cases, especially considering the near-saturated performance on simpler benchmarks like code generation. However, creating a benchmark dataset for migration tasks poses significant challenges, given the breadth and depth of repository-wide changes involved. 

\dblue{To fill in this gap},
we introduce \migrationbench, the first large-scale benchmark dataset in \java, designed specifically to assess LLM capabilities in code migration.
Curating a high-quality migration dataset is a complex and multifaceted task that demands meticulous attention to various factors.
Our approach employs a rigorous and scalable multi-step process that takes into account diverse aspects of the repository. 
We begin by applying license and repository quality filters to ensure compliance and maintain a certain level of quality. 
Subsequently, we employ robust techniques to identify the most appropriate snapshot as the starting point,
ensuring 
the dataset's integrity and reliability.
Furthermore,
we design a comprehensive evaluation framework for migration success
\dblue{due to the absence of ground truth}.
Our approach ensures that
\dblue{both the benchmark datasets and final evaluation} are of exceptional quality,
providing researchers and practitioners with a reliable and comprehensive resource for studying and analyzing code migration patterns.



Our major contributions are summarized below:
\begin{enumerate}
    \item We introduce \migrationbench\ for repository-level code migration task from \javan{8},
    \textit{the} most popular programming language (PL) for around $20$ years\footnoteurl{https://www.tiobe.com/tiobe-index},
    the \textit{first} large scale benchmark dataset
    among all PLs.
    The full MigrationBench dataset (\cmd{full}) contains 5{,}102 open-source Maven-based Java~8 repositories collected from GitHub. To accelerate future research on code migration, we further curate a more representative and challenging subset, \lite, consisting of 300 repositories sampled from \cmd{full}. Both datasets are publicly hosted at \url{https://huggingface.co/collections/AmazonScience}.

    \item 
    We propose an automated and comprehensive evaluation framework that verifies minimal migration requirements, \textit{approximately} validates functional equivalence (FE) after migration, and checks whether dependencies are upgraded to their \textit{latest} major versions in \url{\githubUrl}.

    \item We present an agentic workflow built on the Strands agent framework\footnoteurl{https://strandsagents.com} to address code migration tasks.
    By providing basic shell access and migration knowledge base, the agent achieves reasonable migration efficacy.
    To further improve efficiency, we propose a hybrid approach that combines static code analysis with LLM-based agents. This hybrid method achieves performance comparable to the purely agentic approach while reducing LLM usage by 11\%. {We release source code at \url{https://github.com/amazon-science/JavaMigration}.}

    \item
    {Together with the source code, we also} release all agent trajectories and their evaluation for \lite\ subset to support reproducibility and facilitate future research. These trajectories provide fine-grained records of agent behavior, including intermediate decisions, tool invocations, and failure modes, which are critical for understanding the limitations of current agentic approaches. Beyond analysis, the released trajectories can also serve as valuable training data for developing and improving learning-based methods for code migration.

\end{enumerate}

The paper is structured as follows:
\cref{sec:related-work}
provides an overview of related work in the field, situating our research within the existing body of knowledge.
\cref{sec:migration-bench}
details our meticulous process to curate the dataset.
In \cref{sec:eval},
we delve into the evaluation metrics.
\cref{sec:agent}
introduces our baseline solution with strands agent, an enhanced version by prompt engineering, a RAG-agent with migration knowledge base, and a hybrid approach combining static code analysis and LLM-based agents.
\cref{sec:exp} presents the experimental results obtained from applying our agentic solution to the curated dataset. Finally,
we summarize 
major findings in \cref{sec:conclusion}.
\section{Related Work}
\label{sec:related-work}


\textbf{LLM Applications in the Software Engineering Industry.}
In the software engineering (SWE) industry, code maintenance~\cite{canfora2001software} such as platform migration,
library version upgrades, code refactoring and bug fixing, is an essential component in daily SWE work, requiring even significantly more SWE hours than developing new features~\cite{lientz1978characteristics}.
Therefore, the community has an increasing interest in leveraging LLMs in nearly all fields of SWE tasks~\cite{ding2024reasoning}. 
\dblue{Reflexion \cite{shinn2024reflexion},
Agentless \cite{xia2024agentless} and 
Agentcoder \cite{huang2023agentcoder}
}
mainly study the LLMs' ability in code generation to solve the interview-style coding problems.
\dblue{ChatUniTest \cite{chen2024chatunitest}
and \textsc{TestPilot} \cite{schafer2023empirical}
} explore the feasibility to use LLMs for automated unit test generation (UTG),
\dblue{while Toggle investigates} fault localization and program repair \cite{Hossain2024}.


Building on these foundational applications, the introduction of workflows and agents further enhances the utility of LLMs by addressing key limitations like hallucination and reasoning constraints through external feedback mechanisms~\cite{hong2023metagpt, huang2023agentcoder, yoffe2024debunc}. Tools such as OpenDevin~\cite{openhands} and Cursor\footnoteurl{https://www.cursor.com} excel at resolving complex coding tasks, automating repetitive processes
and offering intelligent debugging solutions. Similarly, GitHub Copilot\footnoteurl{https://github.com/features/copilot} and Amazon Q Developer\footnoteurl{https://aws.amazon.com/q/developer} stand out for providing real-time code suggestions and enhancing developer productivity, while Replit Ghostwriter\footnoteurl{https://replit.com/learn/intro-to-ghostwriter} focuses on collaborative coding and debugging. These innovations showcase how workflows and agents enable LLMs to handle longer, more complex trajectories effectively.

\textbf{LLM Agents for Software Engineering.}
Recent research has increasingly investigated large language model (LLM)–based agents for solving software engineering (SWE) tasks, particularly in the context of automated bug fixing, code repair, and issue resolution. Many recent works propose agentic frameworks that decompose SWE tasks into multi-step decision processes, enabling LLMs to iteratively plan, invoke tools (e.g., search, execution, and testing), and incorporate feedback from the environment \cite{yao2022react, shinn2024reflexion, zhang2024codeagent, yang2024swe, xia2025live}. More recent methods explore search-based and reinforcement-learning–inspired strategies to guide action selection and refinement \cite{jin2025search}. Despite these advances, there has been no systematic evaluation of LLMs on the code migration task. To address this gap, we propose an agentic framework for code migration on \migrationbench.

\textbf{Benchmark Datasets: Code Generation, Maintenance and Translation.}
Benchmark datasets for coding tasks are abundant,
but most focus on use cases other than code migration,
\dblue{as shown in a survey of around $200$ coding benchmark datasets
covering all phases of the software development life cycle comprehensively
\cite{wang2025sdlc}}.
HumanEval~\cite{humaneval} is a coding benchmark dataset designed to evaluate the correctness of LLMs in generating \cmd{Python} functions based on the provided docstrings. Mostly Basic Programming Problems (MBPP)~\cite{mbpp} offers a more diverse dataset, assessing LLMs' overall problem-solving abilities in real-world coding scenarios with natural language problem descriptions. These datasets have been further enhanced into HumanEval$^+$ and MBPP$^+$~\cite{liu2024your} to provide a more comprehensive evaluation framework for LLMs. APPS~\cite{hendrycks2021measuring} extends the prior work by measuring the model's ability to take an arbitrary natural language specification and generate \cmd{Python} code.
While the aforementioned benchmarks primarily focus on \cmd{Python} code,
MBXP and Multilingual HumanEval~\cite{athiwaratkun2022multi} extend the evaluation to over $10$ PLs,
enabling a more comprehensive assessment of multilingual code generation capabilities.

\dblue{DevAI \cite{zhuge2024agent}, a benchmark of 55 realistic AI development tasks organized hierarchically as a directed acyclic graph (DAG), evaluates agentic systems on their ability to handle diverse aspects of AI development pipelines.}
\dblue{In code maintenance,}
SWE-Bench 
\cite{jimenez2024swebenchlanguagemodelsresolve}
compiles a collection of popular \cmd{Python} repositories to facilitate research on leveraging LLMs for effective \github\ issue resolution.
{
While SWE-Bench-Java \cite{zan2024swebenchjava} expands to the \cmd{Java} PL with $91$ issues,
SWE-PolyBench \cite{rashid2025swepolybench}
is more comprehensive
with $2,110$ issues in multiple PLs, 
including \cmd{Java}, \cmd{JavaScript}, \cmd{TypeScript} and \cmd{Python}.}
Notably, while DevAI, SWE-Bench and its extensions include components that
address broader development contexts
\dblue{with more advanced and difficult SWE tasks},
none of them comprehensively tackles coding problems at the \textit{repository} level as \migrationbench\ does.

Instead, repository-level research has been primarily focusing on code translation tasks, typically involving a few medium-sized \repos.
Therefore, it lacks both a well-designed comprehensive benchmark dataset and sufficient complexity.
While AlphaTrans \cite{ibrahimzada2024repoalphatrans} focuses on translating \java\ to \cmd{Python} for $10$ 
\repos\ and RepoTransBench \cite{wang2024repotransbench} attempts to translate $100$ \repos\ 
in the opposite direction,
neither of them has a ground truth target \repo\ available.
There are more research on translation to \cmd{Rust} as the target PL,
from a source PL of \cmd{Go} \cite{zhang2024scalablecodetransgo2rust} or
\cmd{C} \cite{shetty2024syzygydualcodetestc2rust}.
These translations only involve less than $10$ \repos\ and up to $7,000$ lines of code (LOC) for each \repo.
CodePlan \cite{bairi2024codeplan} is closer to our work,
demonstrating LLMs can resolve \cmd{C\#} package migration and
\cmd{Python} code edits,
with $3$ \repos\ for each use case.
Several independent code migration benchmarks have been proposed \textit{after} \migrationbench. \textsc{CodeMenv}~\cite{cheng-etal-2025-codemenv} studies code migration at the \textit{function} level, allowing up to three incompatible lines, and focuses on \cmd{Python} and \cmd{Java}. \textsc{FreshBrew}~\cite{may2025freshbrew} emphasizes test coverage and includes 228 repositories, but adopts a less comprehensive curation and evaluation pipeline.
In contrast, \migrationbench\ is a large-scale benchmark comprising more than 5{,}000 repositories and is accompanied by a robust, automated evaluation framework, making it well suited for systematic and reproducible evaluation of code migration methods.


\section{MigrationBench}
\label{sec:migration-bench}

\href{https://huggingface.co/collections/AmazonScience/migration-bench-68125452fc21a4564b92b6c3}{\migrationbench}
is a benchmark comprising a collection of
\github\ open-source repositories,
all written in a specific source language version ($L_S$).
Given a target language version ($L_T$),
the task is to migrate the repository to ensure compatibility with $L_T$ 
\dblue{with necessary code changes.}
\dblue{In this paper, we present its \java\ version,
that is $L_S = \text{\javan{8}}$ and build tool is \maven.}
These Maven-based applications are intended to be migrated to long-term support (LTS) $L_T$ such as \javan{17} or $21$.
To facilitate large-scale \repo\ processing,
we leverage AWS Elastic Map Reduce (EMR) Serverless,
running under \cmd{Amazon Linux 2}
and \maven\;\cmd{3.9.6},
\dblue{with $4$ vCPUs and $16$G memory for dataset curation}.

\subsection{Problem Formulation}
\label{ssec:prob_form}

{\bf Notations.}
Let $R$ denote the content of a code repository including dependency declaration files (e.g. \cmd{pom.xml}),
source code files,
test files,
auxiliary files (e.g. README, graphs, tables), etc.
$R$ is usually associated with a PL version $L$ (e.g. \javan{8}, $17$, $21$) and can be built, tested or verified by a collection of verifiers $\mathbf{v}$. For example, a Maven \java\ \repo\ $R$ can be built and tested by $\mathbf{v}=\text{\mvnverify}$. 
We define a status function $f(R, L, \mathbf{v})$ that maps a code \repo\ $R$,
its associated language $L$
and verifiers $\mathbf{v}$ to a boolean value,
representing whether the verifiers pass or not, i.e. $f(R, L, \mathbf{v})=\text{\cmd{True}}$ or \cmd{False}. 


Let $R_\dblue{S}$ and $R_\dblue{T}$ denote the code \repo\ before and after the migration respectively. 
As \migrationbench\ does \textit{not} aim at bug fixing,
we require that the original code repository $R_S$ is in a passing state before the migration under source language $L_S$,
i.e. $f(R_S, L_S, \mathbf{v}) = \text{\cmd{True}}$.
Given a target language $L_T$,
we formulate the migration problem as to find a transformation from $R_S$ to $R_T$ such that $f(R_T, L_T, \mathbf{v}) = \text{\cmd{True}}$.
Note that the ending states $R_T$
subject to $f(R_T, L_T, \mathbf{v}) = \text{\cmd{True}}$
are usually not unique.

In this paper we specifically discuss the scenario
where $L_S$ and $L_T$
are \javan{8} and $17$ respectively,
and \mvnverify\ is one component of the 
verifiers $\mathbf{v}$.

Note that some repositories may have hard-coded compiler versions in the dependency files (\cmd{pom.xml}),
then they require \textit{seed changes} (\seedchange) \cite{bairi2024codeplan} when migrating a \repo. For example, a valid seed change can 
replace
any hard-coded \javan{8}\ with $17$\ in \cmd{pom.xml} files.

\subsection{Data Collection}
\label{subsec:data_collection}

We introduce the data collection process which produces the \migrationbench\ dataset. 
We scrape \textit{all} \java\ \repos\ hosted in \github\ and apply multiple filters 
$\mathbf{F} \equiv \{F_i, 1 \leq i \leq 6\}$
sequentially
detailed below to ensure high dataset quality.


\textbf{\filter{1}.\ Licenses.}
We collect Github \repos\ under the \cmd{MIT} or
\cmd{Apache $2.0$} licenses only.


\textbf{\filter{2}. Enforce Repository Quality through Github Stars.}
To minimize the inclusion of potentially low-quality \repos,
we require \repos\ to have at least $3$ stars.

\textbf{\filter{3}.\;Build Tool is \maven.}
For \java\ projects,
Apache Maven,
Gradle
and legacy Apache Ant
are all widely used build automation tools
to manage dependencies, compile packages and run tests.
In this paper, we limit the scope to \maven, which is not only the most popular option for \javan{8} but also represents a mature and well-documented build tool.
As part of our criteria,
a \repo's \textit{latest} snapshot must
pass with the \mvnverify\ command using at least one of \javan{8}, $11$ or $17$ without any modifications.
This ensures that each selected \repo\ is in a stable and passing state,
so that it's reasonable to assume the existence of a valid migration path for those \repos.

By applying filtering up to \filter{3},
we narrow down to $\mathbf{16,154}$ repositories as of March $2024$.
All subsequent filters are conducted 
in March $2025$.

{\bf Note.}
\dblue{
We encounter significant \cmd{maven} throttling when executing a high number of concurrent jobs,
leading to potential \textit{incompleteness} in the $16$k list.
\begin{enumerate}
\item 
Despite this,
we consider false positives to be more detrimental than false negatives,
thus the $16$k list serves as a sufficiently robust and large-scale starting point for benchmark datasets like \migrationbench.
\item
Moreover,
while false negatives may persist up to stage \filter{3},
we have adopted best-effort runtime strategies in subsequent stages,
\filter{4} through \filter{6},
to mitigate the throttling issue.
\end{enumerate}
}



\textbf{\filter{4}. Search for $L_S$-compatible Base Commit ID ($H_b$).}
We check that the \github\ url remains valid,
and then search for the last commit id,
i.e.\ base commit $H_b$,
which is compatible under source language $L_S$ for each \repo.
$L_S$-compatibility for a \repo's given snapshot is defined as below:
\begin{enumerate}
\item It has at least one \cmd{pom.xml} file.
\item It has valid source versions,
i.e.
each \cmd{pom.xml} file in a \repo\ has {either} no hard-coded \java\ versions
or the hard-coded version is \javan{8},
defined either through\\
\emph{maven.compiler.source}, \emph{maven.compiler.target} and\\
\emph{maven.compiler.release} fields in \emph{properties}, or through artifact \emph{maven-compiler-plugin}'s \emph{source}, \emph{target} and \emph{release} configurations.
\item It's able to pass \cmd{mvn clean verify} with \javan{8}
and 
\item We further require the major versions of the final compiled classes,
exported to the \cmd{target/classes/} or customized output directory,
are $52$.
This class validation makes sure the \textit{effective} \cmd{pom.xml} files are indeed \javan{8}.
\end{enumerate}

\cref{app:subsec:base_commit_search_appendix} covers more details on the
assumptions and approximations for a faithful collection.

In the end,
it reduces to $\mathbf{9,934}$ \repos\ with valid $H_b$ and at least one \cmd{*.java} file present.

\textbf{\filter{5}. Dedup Repositories by Exact Match.}
There are numerous duplicate \repos\ on \github\ and \filter{5} aims at removing them.
Given a snapshot for a \repo,
we compute its \textit{snapshot} id $H^S$ for the whole \repo\ at base commit $H_b$,
by hashing the tree structure of the \repo's files,
\textit{each} \cmd{*.java} file's hash
and 
both filenames (relative to root) and
raw content for \textit{each} \cmd{pom.xml} file,
concatenated by new lines.
De-duplicating based on $H^S$
results in $\mathbf{9,916}$ \repos\
and they'll be used as the starting point for \migrationbench.

\textbf{\filter{6}. Partition into \cmd{full} and \cmd{UTG} Subsets.}
As discussed in \cref{subsec:discussion}, the migration efficacy can be significantly different when including unit tests. Therefore, we further require \repos\ to have at least one test case to ensure at least partial
functional equivalence after migration.
We identify unit tests based on:
\begin{enumerate}
\item Whether there are any \cmd{*.java} files present in the dedicated test directory \cmd{src/test/} or
\item Whether there are any test cases triggered by \cmd{mvn test -f .}\ command in the root directory.
\end{enumerate}

We further partition these \repos\ into a \cmd{full} dataset containing $\mathbf{5,102}$ \repos\ with unit tests, 
and a \cmd{UTG} subset containing the remaining $\mathbf{4,814}$ \repos\ without any unit tests.
The remaining of this work primarily discusses \cmd{full} due to its direct relevance to migration evaluation,
and we leave the \cmd{UTG} subset as a standalone benchmark for unit test generation.


{

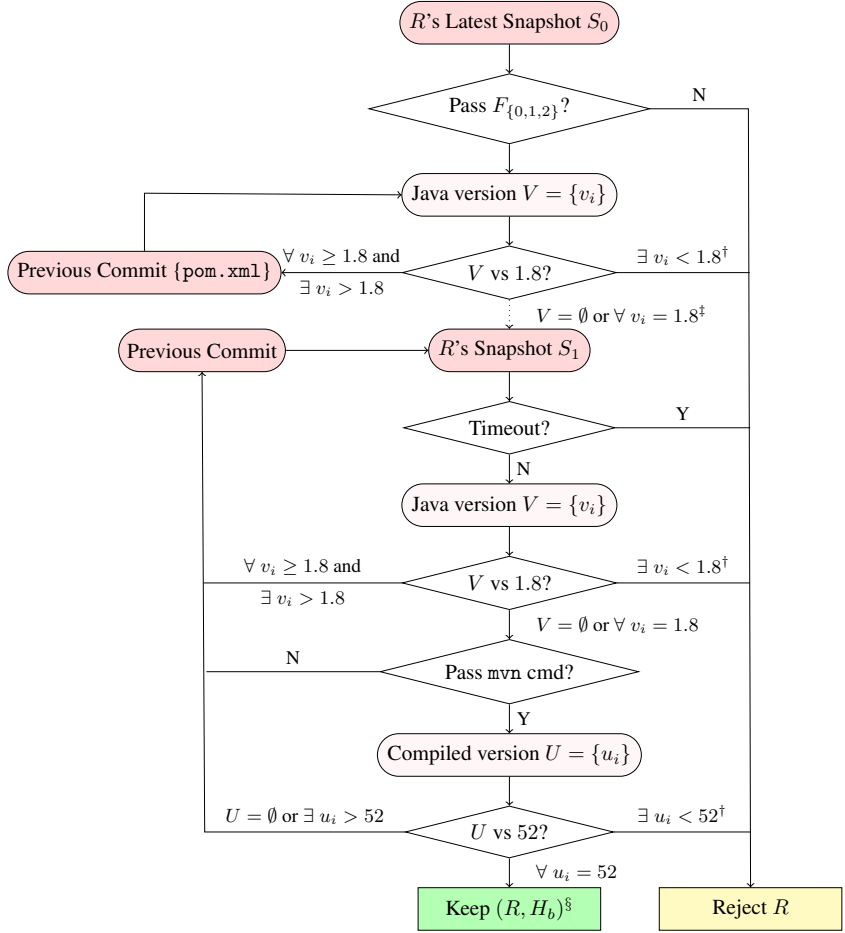
\begin{figure}[!ht]
\resizebox{\columnwidth}{!}
{
\tikzstyle{system} = [draw, 
rectangle, node distance=2em, minimum width=3cm, minimum height=0.7cm]
\tikzstyle{system2} = [draw, fill=green!30, rectangle, node distance=2em, minimum width=2.4cm, minimum height=0.7cm]
\tikzstyle{state} = [draw, fill=red!15, rounded rectangle, node distance=2em, minimum width=2cm, minimum height=0.7cm]
\tikzstyle{statelight} = [draw, fill=red!3, rounded rectangle, node distance=2em, minimum width=3cm, minimum height=0.7cm]
\tikzstyle{decision} = [draw, diamond, aspect=4, node distance=2em]

\begin{tikzpicture}[every text node part/.style={align=center}]
\node [state, name=repos] {$R$'s Latest Snapshot $S_0$};
\node [decision, name=filters, below=3ex of repos] {Pass $F_{\{0, 1, 2\}}$?};
\node [statelight, name=java_version, below=3ex of filters] {Java version $V=\{v_i\}$};

\node [decision, name=java_version_cmp, below=3ex of java_version] {$V$ vs $1.8$?};
\node [state, name=previous_commit_pom, left=10.5ex of java_version_cmp] {Previous Commit\\\{\cmd{pom.xml}\}};
\node [state, name=start_commit, below=3ex of java_version_cmp] {$R$'s Snapshot $S_1$};
\node [decision, name=timeout, below=3ex of start_commit] {Timeout?};
\node [statelight, name=j_version, below=3ex of timeout] {Java version $V = \{v_i\}$};
\node [decision, name=j_version_cmp, below=3ex of j_version] {$V$ vs $1.8$?};
\node [decision, name=mvn_clean_verify, below=3ex of j_version_cmp] {Pass \cmd{mvn} cmd?};
\node [state, name=previous_commit, left=14.8ex of start_commit] {Previous Commit};

\node [statelight, name=c_version, below=3ex of mvn_clean_verify] {Compiled version $U=\{u_i\}$};

\node [decision, name=c_version_cmp, below=3ex of c_version] {$U$ vs $52$?};

\node [system2, name=accept, below=3ex of c_version_cmp] {Keep $(R, H_b)^\S$};
\node [system2, name=reject, fill=yellow!30, minimum width=1.6cm, right=8ex of accept] {Reject $R$};

\draw [->] (repos) -- (filters);

\draw [->] (filters) -- (java_version);
\draw [->] (java_version) -- (java_version_cmp);
\draw [dotted,->] (java_version_cmp) -- (start_commit)
node[pos=0.5,right,font=\footnotesize]
{\quad $V=\emptyset \text{ or } \forall\;v_i=1.8^{\ddagger}$};

\draw [->] (java_version_cmp) -- (previous_commit_pom)
node[pos=0.4,above,font=\footnotesize]
{$\forall\;v_i\geq1.8$ and}
node[pos=0.5,below,font=\footnotesize]{$\exists\;v_i>1.8$};

\draw [->] (start_commit) -- (timeout);
\draw [->] (timeout) -- (j_version)
node[pos=0.5,right,font=\footnotesize]{N};
\draw (timeout.east)
-- ++(11.55ex, 0)
node[pos=0.5,above,font=\footnotesize]{Y};
\draw [->] (j_version) -- (j_version_cmp);
\draw [->] (j_version_cmp) -- (mvn_clean_verify)
node[pos=0.5,right,font=\footnotesize]
{\quad \underlinejv{$V=\emptyset \text{ or } \forall\;v_i=1.8$}};

\draw [->] (previous_commit_pom.north)
-- ++(0, 5.2ex)
-- (java_version.west);

\draw [->] (previous_commit.east)
-- (start_commit.west);

\draw (mvn_clean_verify.west) 
-- ++(-18.2ex, 0)
node[pos=0.5,above,font=\footnotesize]{N};
\draw [->] (mvn_clean_verify) -- (c_version)
node[pos=0.5,right,font=\footnotesize]{Y};
\draw [->] (c_version) -- (c_version_cmp);
\draw (j_version_cmp.west)
-- ++(-21ex, 0)
node[pos=0.5,above,font=\footnotesize]
{\underlinejv{$\forall\;v_i\geq1.8$ and}}
node[pos=0.5,below,font=\footnotesize]{\underlinejv{$\exists\;v_i>1.8$}};

\draw [->] (filters.east)
-- ++(8.2ex, 0)
node[pos=0.5,above,font=\footnotesize]{N}
-- (reject.north);
\draw (java_version_cmp.east)
-- ++(11.7ex, 0)
node[pos=0.5,above,font=\footnotesize]{$\exists\; v_i<1.8^\dagger$};
\draw (j_version_cmp.east)
-- ++(11.6ex, 0)
node[pos=0.5,above,font=\footnotesize]{\underlinejv{$\exists\; v_i<1.8^{\dagger}$}};
\draw (c_version_cmp.east)
-- ++(11.8ex, 0)
node[pos=0.5,above,font=\footnotesize]{$\exists\; u_i<52^\dagger$};
\draw [->] (c_version_cmp.west)
-- ++(-21.2ex, 0)
node[pos=0.5,above,font=\footnotesize]{$U=\emptyset$ or $\exists\;u_i>52$}
-- (previous_commit.south);
\draw [->] (c_version_cmp) -- (accept)
node[pos=0.5,right,font=\footnotesize]
{\quad $\forall\;u_i=52$};
\end{tikzpicture}
}
\caption{The flowchart to find out the base commit id $H_b$ for code migration,
starting with a \github\ \repo\ $R$'s latest snapshot as of March $2024$.}
\label{fig:data_workflow}
\Description{A flowchart to find out a Github repository's historical commit id from a given snapshot.}

\parbox{\columnwidth}{\footnotesize
$^\dagger$ It assumes both the \textit{explicitly} hard coded \cmd{java} versions and 
complied
class major versions are monotonic over time.\\
$^\ddagger$ Given the commit history of $V$
based on all \cmd{pom.xml} files,
one can infer the first candidate snapshot $S_1$.\\
$^\S$ $H_b$ is the commit id of the \textit{last} $S_1$ where
java version $V$ passes version check,
\mvn\ command succeeds and
compiled class major version $U$ passes version check.
}

\end{figure}

\subsection{\href{https://huggingface.co/datasets/AmazonScience/migration-bench-java-selected}{\migrationlite\ (\lite)}: A Representative and Challenging Subset}
While we collect thousands of open-source repositories from \github\ to cover the \textit{entire} problem space,
the whole dataset is not always the best benchmark for evaluating LLMs' migration capabilities.
\dblue{A significant portion of the repositories may share similar issues},
whereas a benchmark dataset should be diverse enough to reflect a range of challenges encountered in real-world migrations.
Additionally, running LLM-assisted migration 
on all $5$k repositories is prohibitively time- and resource-intensive.

{We further introduce a smaller dataset to address these limitations, \href{https://huggingface.co/datasets/AmazonScience/migration-bench-java-selected}{\migrationlite\ (\lite)}.}
\lite\ is a subset of $\mathbf{300}$ \repos\ 
carefully curated from \cmd{full}
based on objective \repo\ statistics and through human expert selection, which place greater emphasis on larger and multi-module repositories.
\dblue{
By prioritizing \repos\ that present more complex migration challenges,
\lite\ is designed to better capture the diversity and intricacies of real-world code migration scenarios.
}

\subsection{Dataset Statistics}
\label{subsec:migration-bench-stats}

We report statistics for both \migrationbench's \cmd{full} and \lite\ subsets in \cref{tab:dataset_stats}.
On average,
\repos\ in the \cmd{full}\ subset have $2.4$ modules,
$7,698.3$ cumulative LOC,
$65.1$ \cmd{java} files,
$8.7$ test files in the \cmd{src/test} directory
and \dblue{$34.8$ test cases}.
For the \lite\ subset,
metrics' average, median and standard deviation values are typically larger than the \cmd{full} superset.
\begin{table*}[!ht]
\centering
    \caption{\migrationbench\ statistics at \cmd{base commit} $H_b$ for both the \cmd{full} and the \lite\ subset.
For rows reporting average values, the highest and second-highest maximal values are highlighted in bold and underlined, respectively.
}
    \label{tab:dataset_stats}
    \scalebox{1}{
    \begin{tabular}{llrrrr}
    \toprule
    \multirow{2}{*}{Metric$^\dagger$}
    & \multirow{2}{*}{Statistics}
    & \multirow{2}{*}{\cmd{full}$\hidezero$}
    & \multicolumn{3}{c}{\lite}
    \\
    \cmidrule{4-6}
    \\
    &&
    & {all$\hidezero$}
    & {success$^\ddagger\hidezero$}
    & {failure$^\ddagger\hidezero$}
    \\
    \midrule  
    Size &
        -
        & $5,102\hidezero$
        & $300\hidezero$ 
        & $46\hidezero$
        & $254\hidezero$ \\
    \cmidrule{1-1}
    \multirow{4}{*}{\#Modules} 
        & \cmd{avg}     & $2.4$          &  $\underline{6.6}$ & $2.4$ & $\mathbf{7.4}$ \\
        & \cmd{std}     & $5.8$          & $\mathbf{15.1}$ &- &- \\
        & \cmd{median}  &   $1\hidezero$ &   $1\hidezero$ &- &- \\
        & \cmd{max}     & $192\hidezero$ & $192\hidezero$ &- &-\\
    \cmidrule{1-1}
    \multirow{4}{*}{LOC}
        & \cmd{avg} & $7,698.3$ & $\underline{22,397.2}$ & $\mathbf{23,771.2}$ & ${22,148.4}$ \\
        & \cmd{std} & $35,937.3$ & $\mathbf{108,203.9}$  &- &-\\
        & \cmd{median} & $2,044.5$ & $\mathbf{7,351.5}$ &- &- \\
        & \cmd{max} & $ 1,773,940\hidezero$ & $1,773,940\hidezero$ &- &-\\
    \cmidrule{1-1}
    \multirow{4}{*}{\#\java\ Files}
        & \cmd{avg}     &  $65.1$           & $\underline{168.4}$ & $123.4$ & $\mathbf{176.5}$\\
        & \cmd{std}     & $218.6$           & $\mathbf{650.7}$ &- &- \\
        & \cmd{median}  & $25\hidezero$      & $\mathbf{82\hidezero}$ &- &-\\
        & \cmd{max}     & $10,970\hidezero$ & $10,970\hidezero$ &- &- \\
    \cmidrule{1-1}
    \multirow{4}{*}{\#Test Files$^\S$}
        & \cmd{avg}     &  $8.7$        & $\underline{8.8}$ & $\mathbf{9.3}$ & $8.7$ \\
        & \cmd{std}     & $\mathbf{26.6}$        & $21.8$  &- &-\\
        & \cmd{median}  &  $\mathbf{2\hidezero}$ & $1\hidezero$ &- &- \\
        & \cmd{max}     & $\mathbf{815\hidezero}$ & $194\hidezero$ &- &- \\
    \cmidrule{1-1}
    \multirow{4}{*}{\dblue{\#Test Cases$^\star$}}
        & \cmd{avg}     &  $34.8$        & $\underline{110.6}$ & $\mathbf{354.9}$ & ${70.0}$ \\
        & \cmd{std}     & $289.9$        & $\mathbf{854.1}$  &- &-\\
        & \cmd{median}  &  $2\hidezero$ & $\mathbf{8.5}$ &- &- \\
        & \cmd{max}     & $14,450\hidezero$ & $14,450\hidezero$ &- &- \\
    \bottomrule 
    \end{tabular}
}   
\parbox{0.6\textwidth}{\footnotesize
    $^\dagger$ Other than the Size metric,
    all other metrics are computed on a per-\repo\ basis.\\
    $^\ddagger$ We partition the whole \cmd{selected} dataset (\cmd{all})
    into \cmd{success} and \cmd{failure} subsets,
    based on the maximal migration result of baseline Strands agent with \claude.\\
    $^\S$ It counts all \cmd{*.java} files in the \cmd{src/test/} directory only.\\
    $^\star$ It counts test cases by executing \cmd{mvn test -f .} in the root directory.
}
\end{table*}

\section{Evaluation}
\label{sec:eval}
\dblue{
Measuring the success of typical code generation or SWE benchmarking tasks is usually well-defined,
as either the ground truth solution is available
or 
it's straightforward to conduct an evaluation with comprehensive tests focusing on a limited scope e.g. with only \textit{function} level or \textit{file} level changes.}
Thus, they have very limited room for variations though the solutions are not necessarily unique.
However,
none of them applies for coding problems at the \textit{\repo} level.
{Furthermore, code}
migration evaluation presents unique challenges
in that
necessary code changes to $L_T$ typically scale with repository size,
which may vary significantly depending on the application types.
}

\textbf{Approximate Functional Equivalence.}
The most rigorous way to evaluate success is to verify FE before and after migration,
which is challenging by nature due to the huge problem and solution space.
Instead,
we adopt a few key metrics as approximations below,
primarily replying on \mvnverify\ or \cmd{compile} command
and a few {\bf invariants}
as the measure of migration success,
to automate the final evaluation and
enable comparison
across various migration efforts.

\subsection{\dblue{Minimal Migration}}
\label{subsec:migration-req}
\label{subsec:min-migration}

\textbf{Maven Command Success ($r_1$).}
For a successful migration to \javan{17},
the minimum requirement is that the application builds successfully and passes all existing unit and integration tests when running on \javan{17},
verified by the \mvnverify\ command.
It performs a series of steps:
it cleans cached files,
compiles the source code,
runs unit tests,
packages the application,
executes integration tests
and validates the entire build process.
When it completes without any build or test errors,
the build is considered successful. 

\textbf{Compiled Class Major Version Validation ($r_2$).}
To ensure that the compiler is genuinely using \javan{17},
we implement a guardrail to verify the compiled classes' major versions
are indeed $61$.\footnoteurl{https://mkyong.com/java/list-of-java-class-file-major-version-numbers/} These compiled classes are stored in \cmd{*/target/classes/**/*.class} usually.

\textbf{Invariance: List of Test Methods ($r_3$).}
To make sure test methods (annotated with \cmd{@Test}) are 
not renamed{, disabled} or even {completely} removed after code migration,
we compare test classes and annotated test methods based on abstract syntax tree (AST) parsing for all files in the dedicated \cmd{src/test} directory before and after migration,
and the migration is considered unsuccessful if the test method names don't match.

\textbf{Invariance: Number of Test Cases ($r_4$).}
Static analysis of test directory in $r_3$ still does {\bf not} guarantee runtime behaviors,
therefore we further require number of test cases to be non-decreasing,
primarily to avoid disabling those tests in \cmd{pom.xml} files
hence effectively treating them as plain text files.

When a migrated \repo\ satisfies requirements $\{r_1, r_2, {r_3, r_4}\}$,
we define it as \textbf{minimal migration} and the corresponding efficacy (\cmd{pass@$1$}) is:

\begin{equation}
\eta_{\etamin} = \displaystyle\frac
{\scriptsize \text{\#\ migrated \repos\ passing $\{r_1,r_2, r_3, r_4\}$}}
{\scriptsize \text{\#\ total \repos}}
\end{equation}

\subsection{\dblue{Maximal Migration}}
\label{subsec:max-migration}

\textbf{{Dependencies' \textit{Latest} Major Versions ($r_5$).}}
One key advantage of upgrading \java\ is 
to leverage
the dependency libraries' latest  major versions for efficiency and security reasons.
For instance, the latest Spring Boot version $3.3.\text{x}$ is incompatible with \javan{8},
requiring at least \javan{17}
since $3.0$.\footnoteurl{https://docs.spring.io/spring-boot/docs/3.0.0/reference/html/getting-started.html}
Therefore,
to ensure a modern and robust migration,
\dblue{
we require using \textbf{all} dependencies' stable and
\textit{latest}
{major} versions available on Maven Central
as of November $2024$.
}
We specify a list of version requirements for the most frequent $240$ dependencies occurring in the \lite\ subset. For the complete list of dependency versions, check \url{https://github.com/amazon-science/MigrationBench/blob/main/src/migration_bench/reference/dependency_version.json}.

Similar to minimal migration,
we define \textbf{maximal migration} and the corresponding efficacy is:

\begin{equation}
\eta_{\etamax} = \displaystyle\frac
{\footnotesize \text{\#\ migrated \repos\ passing $\{r_1,r_2, r_3, r_4, \mathbf{r_5}\}$}}
{\footnotesize \text{\#\ total \repos}}
\end{equation}

Note that (i) By definition
$\eta_\etamax \leq \eta_\etamin$ and
(ii) We leverage \cmd{mvn dependency:tree} command to retrieve the effective version for each dependency, instead of simply relying on the dependency declaration files (\cmd{pom.xml})


\textbf{Compilation Efficacy.}
We also define $ \eta_\etamin^{\compile}$ and $\eta_\etamax^{\compile}$,
by
\dblue{dropping $\{r_3, r_4\}$ and
replacing $r_1$ with
$r_1^{\prime}$,
a \textit{weaker} command \cmd{mvn clean compile},
to 
address compilation errors only.\footnoteurl{https://maven.apache.org/guides/introduction/introduction-to-the-lifecycle}}

Complementary to \cref{subsec:migration-req,subsec:max-migration},
there could be even more specifications on migration tasks,
and an incomplete list
includes
a more comprehensive FE check through
enhancing test coverage in \cref{subsec:appendix_eval_req_ut},
adopting modernized syntax and new security features in $L_T$,
etc.


\section{Agentic Framework}
\label{sec:agent}




\subsection{Strands Agent: Baseline}
Large Language Model (LLM) agents have recently demonstrated strong capabilities in solving software engineering (SWE) tasks, including repository-level code understanding, modification, and validation. Prior work such as \cite{yang2024swe} and \cite{xia2025live} shows that LLM-driven agents can autonomously perform non-trivial development workflows by iteratively reasoning, editing code, and executing commands. These results motivate the use of agentic systems for large-scale Java \repo\ migration tasks.

We adopt the {Strands agents} framework as the backbone of our agentic design. Strands agent is a simple yet powerful SDK for building and running AI agents using a model-driven approach. It supports a wide spectrum of use cases—from lightweight conversational assistants to complex autonomous workflows—and scales seamlessly from local development to production deployment. As a baseline, we equip a Strands agent with a shell tool and an edit tool, enabling it to inspect the repository, modify source files, and execute build commands. The agent is instructed to ensure that \mvnverify\ completes successfully without errors under \javan{17}. This configuration allows the agent to perform basic Java version migration and resolve straightforward compilation or test failures.

\subsection{PE Agent (Baseline + PE)}
While effective for minimal migration,
{the standalone Strands} agent does not {effectively conduct} \emph{maximal migration},
{where all dependencies are upgraded} to their latest major versions 
{as shown in \cref{tab:efficay_main}}.
To address this limitation, we further augment the agent's \textit{system} prompt with explicit instructions to update each dependency to its latest major release whenever possible.
This variant relies purely on prompt engineering (PE) on top of the baseline agent and serves as a stronger baseline for maximal modernization.

\subsection{RAG Agent (Baseline + PE + RAG)}
However, without external knowledge of current dependency versions, the agent may fail to identify the correct latest releases,
especially for fast-evolving or less common libraries.
To overcome this, we construct a dependency knowledge base and enable retrieval-augmented generation (RAG).
In this setting, the agent can query the knowledge base to retrieve authoritative information about the latest major versions of dependencies before performing updates.
We refer to this configuration as the \emph{RAG-agent}.

\subsection{Hybrid Approach}
RAG-agent consumes additional tokens when searching from the knowledge base. To further reduce agent execution cost and improve efficiency, we propose a hybrid approach that combines static code analysis with agentic reasoning.
In this design, a static analysis tool first identifies and upgrades all dependencies to their latest major versions in a deterministic manner. 
The PE agent is then invoked to fix resulting build failures, resolve incompatibilities,
and complete any remaining migration steps.
This hybrid approach leverages the strengths of both {domain knowledge injection through} automated code analysis tooling
and LLM-based agents,
achieving effective maximal migration with lower overall cost.
Check \cref{sec:agent_appendix} for detailed prompt templates.


\section{Agent Prompt Templates}
\label{sec:agent_appendix}

We provide additional details of the agent implementation introduced in \cref{sec:agent}. For the baseline agent, we grant access to a shell tool and an edit tool, enabling it to inspect, modify, and execute code. To restrict the agent’s access to the target repository, the shell tool is sandboxed to operate only within the given project directory. We set the maximum number of agent turns to 80 to balance effectiveness and cost.

\subsection{{System Prompt}}
We show {agent specific} system prompts in this section.


\subsubsection{Strands Agent: Baseline}
Here is the system prompt for the baseline Strands agent:

\begin{tcolorbox}[
  colback=gray!5,
  colframe=gray!40,
  title={System Prompt for the Strands Agent: Baseline},
  fonttitle=\bfseries,
]
\ttfamily\small
You are an expert Java developer assistant who can migrate Java projects from JDK 8 to JDK 17. Make sure \verb|`mvn clean verify`| pass with JDK 17 after migration. When \verb|`mvn clean verify`| succeeds, you can conclude the task. You don't have to provide any summary.
\end{tcolorbox}

\subsubsection{PE Agent}
To emphasize maximal migration,
{for the PE agent} we augment the system prompt with additional instructions requiring all dependencies to be upgraded to their latest major versions.
This prompt engineering explicitly guides the agent to prioritize comprehensive dependency modernization beyond minimal build success{:}

\begin{tcolorbox}[
  colback=gray!5,
  colframe=gray!40,
  title={System Prompt for {the PE Agent}},
  fonttitle=\bfseries,
]
\ttfamily\small
You are an expert Java developer assistant who can migrate Java projects from JDK 8 to JDK 17. Make sure \verb|`mvn clean verify`| pass with JDK 17 after migration. You should update all dependencies in the \verb|`pom.xml`| file to their latest versions that support Java 17. When \verb|`mvn clean verify`| succeeds, you can conclude the task. You don't have to provide any summary.
\end{tcolorbox}

\subsubsection{RAG Agent}
In some cases, the {PE} agent struggles to identify the latest major versions of dependencies using its internal knowledge alone,
particularly when those versions were released after the LLM’s knowledge cutoff.
To address this limitation,
{the RAG agent}
leverages a search tool to retrieve up-to-date dependency information from an external knowledge base{:}

\begin{tcolorbox}[
  colback=gray!5,
  colframe=gray!40,
  title={System Prompt for {the RAG Agent}},
  fonttitle=\bfseries,
]
\ttfamily\small
You are an expert Java developer assistant who can migrate Java projects from JDK 8 to JDK 17.\\
Make sure \verb|`mvn clean verify`| pass with JDK 17 after migration.\\
\\
You have access to a dependency version lookup tool. When updating dependencies in pom.xml:\\
1. Use the search\_dependency\_version tool to look up the recommended Java 17 compatible version for each
dependency\\
2. If a dependency is not found in the database, use your knowledge to select an appropriate version\\
3. Update all dependencies to their Java 17 compatible versions
\end{tcolorbox}

\subsubsection{Hybrid Approach}
To further improve migration efficiency,
{the hybrid approach} combines static code analysis with an agentic method.
Specifically, a static parser first upgrades all dependencies to their latest major versions using version information stored in our knowledge base. While this step is deterministic and cost-effective, it may introduce build or test failures due to breaking changes. We therefore invoke the agent afterward to diagnose and fix the resulting errors and to complete any remaining migration steps.
The system prompt for the hybrid approach is shown below{:}

\begin{tcolorbox}[
  colback=gray!5,
  colframe=gray!40,
  title={System Prompt for {the Hybrid Approach}},
  fonttitle=\bfseries,
]
\ttfamily\small
You are an expert Java developer assistant who can migrate Java projects from JDK 8 to JDK 17. Make sure \verb|`mvn clean verify`| pass with JDK 17 after migration. Dependencies in the \verb|`pom.xml`| file have been updated to their latest versions that support Java 17, but these changes might introduce compatibility issues in the codebase. Please fix any such issues in your migration. Do not downgrade the dependency versions back to their JDK 8 compatible versions.
\end{tcolorbox}

\subsection{{Initial User Prompt}}
For all four agent variants, we use the same {initial} user prompt as the input to ensure a controlled and fair comparison.

\begin{tcolorbox}[
  colback=gray!5,
  colframe=gray!40,
  title={{Initial} User Prompt for All Agents},
  fonttitle=\bfseries,
]
\ttfamily\small
The code repository located at \{repo\_path\} is currently written in Java 8. Please migrate the entire codebase to Java 17.
\end{tcolorbox}

\section{Experiments}
\label{sec:exp}
\begin{table*}[!ht]
\caption{Maximal migration efficacy for \migrationlite. Since PE/ RAG agents and the hybrid approach aim at solving maximal migration problem,
they are not evaluated on minimal migration.}
\label{tab:efficay_main}
\begin{center}
\scalebox{1}{
\begin{tabular}{llccc}
\toprule
Method & Type & $\eta_{\etamax}$ (\%) $\uparrow$ & $\eta_{\etamin}$ (\%) $\uparrow$ & Avg. \# LLM Calls per Repo $\downarrow$\\
\midrule
OpenRewrite        & Static   &  2.00 & 16.33 & -\\
Strands Agent       & Agentic &  15.33 & \textbf{71.67} & \textbf{33.68}\\
 ~~~~~ + PE              & Agentic & 45.67 & - & 49.22\\
 ~~~~~ + PE + RAG            & Agentic & \textbf{53.33} & - & 59.22\\
Hybrid Approach           & Hybrid & \textbf{53.33} & - & 52.55\\
\bottomrule
\end{tabular}
}
\end{center}
\end{table*}

\begin{table}[!ht]
\caption{Maximal migration efficacy for \migrationlite\ with hybrid approach and different models.}
\label{tab:os}
\begin{center}
\scalebox{1}{
\begin{tabular}{llcc}
\toprule
Models  & $\eta_{\etamax}$ (\%) & Avg. \# LLM Calls per Repo\\
\midrule
DeepSeek-V3.1  & 6.33 & \textbf{45.93}\\
Qwen3-Coder-480B  & 22.33 & 50.27\\
GLM-5  & {45.33} & 47.20\\
Claude-4.5-Sonnet  & \textbf{53.33} & 52.55\\
\bottomrule
\end{tabular}
}
\end{center}
\end{table}

\begin{figure}[!ht]
\begin{centering}
\includegraphics[width=\columnwidth]{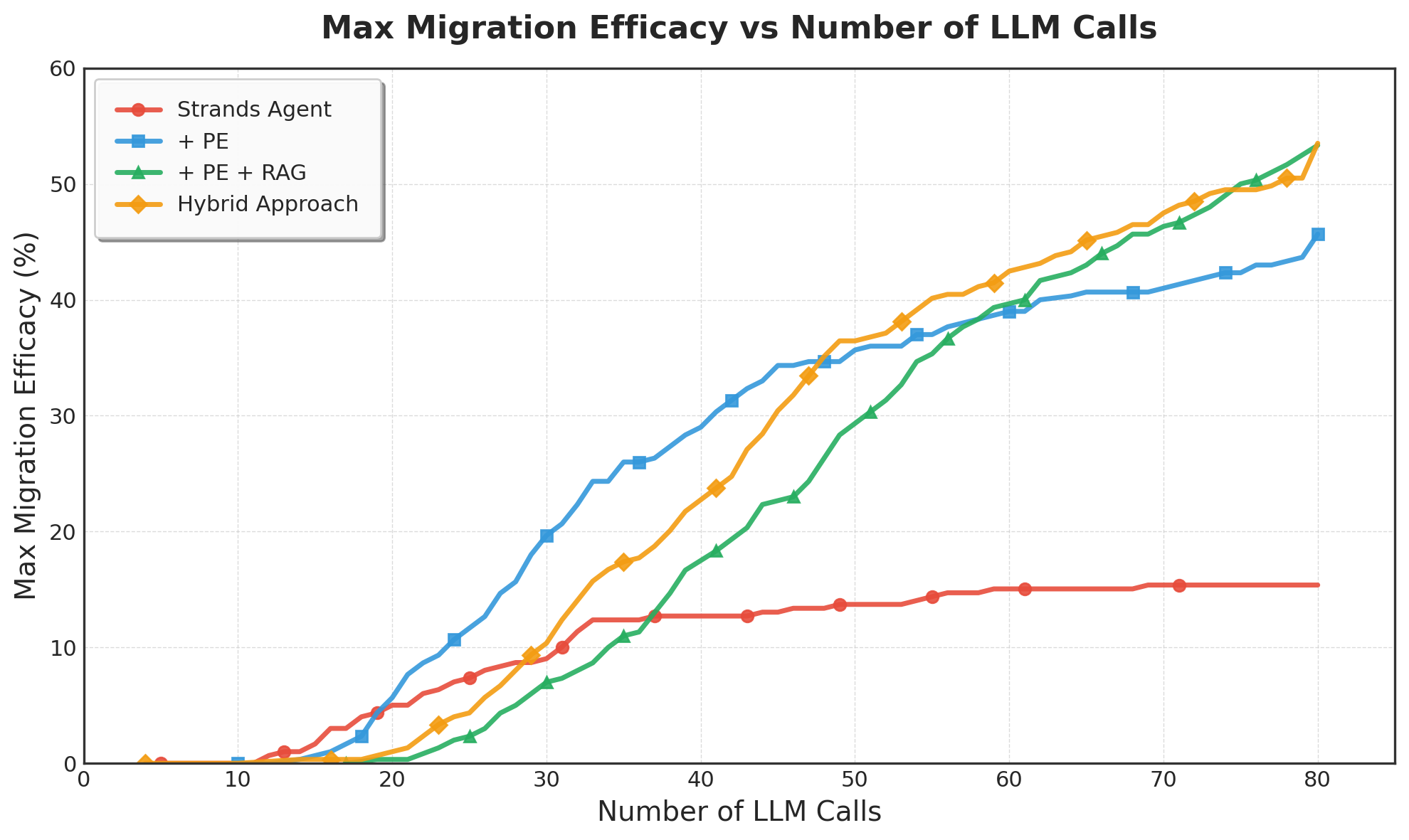}
\caption{
{Maximal} migration efficacy {increases with more} LLM calls for the following methods:
{Strands Agent (standalone baseline) with prompt engineering (+ PE) and RAG (+ PE + RAG), and Hybrid Approach.}}
\label{fig:efficacy}
\Description{A comparison of maximal migration efficacy from a few variations of the Strands agent.}
\end{centering}
\end{figure}

We conduct multiple experiments to collect efficacy metrics as reference points for future research under \lite\ subset.
In addition to the Strands agents and hybrid approach introduced in \cref{sec:agent}, we also consider a static baseline using OpenRewrite recipe \cmd{org.openrewrite.java.migrate.UpgradeToJava17}\footnoteurl{https://docs.openrewrite.org/recipes/java/migrate/upgradetojava17}, which deterministically upgrades Java applications to \javan{17} without involving generative AI.

For all agentic methods, we use {the Strands agent framework with} \claude (thinking-mode enabled)~\cite{claude_4_5_model_card} as the underlying LLM, served via AWS Bedrock. This setup ensures a consistent agent architecture and model provider across experiments.
We release all agent trajectories and their evaluation for \lite\ subset to support reproducibility and facilitate future research.

We also discuss efficacy under two verification criteria,
\mvnverify\ and \mvncompile\, to explicitly highlight the role of unit tests in code migration.
While \mvncompile\ captures basic compilation success,
\mvnverify\ provides a {more strict evaluation} by validating test correctness,
thereby offering a more comprehensive measure of real-world migration quality.

\subsection{Results}

As discussed in \cref{subsec:max-migration}, maximal migration requires more than simply building the \repo\ without errors, which is sufficient only for minimal migration.
{Therefore,} we primarily focus on analyzing maximal migration efficacy,
as it better reflects real-world migration quality~\cite{omidvar2024evaluating}. For completeness, we also report and discuss selected results on minimal migration.

\subsubsection{Efficacy.}
In \cref{tab:efficay_main}, we report minimal and maximal migration efficacy, with a cutoff of ${80}$ LLM API calls (equivalent to 80 turns with the Strands agent) for agentic methods.
Since
{the PE agent, RAG agent and the hybrid approach}
aim at solving maximal migration problem,
they are not evaluated on minimal migration.
We also plot efficacy dependence on the number of LLM API calls in \cref{fig:efficacy}, emphasizing that efficacy is \textit{not} a static number, and it depends on inference-time compute.

\subsubsection{Open-source models.} We also present maximal migration efficacy report for open source models such as \cmd{DeepSeek-V3.1}~\cite{liu2024deepseek}, \cmd{Qwen3-Coder-480B}~\cite{yang2025qwen3} and \cmd{GLM-5}~\cite{zeng2026glm}. We run the hybrid approach on \migrationlite\ with the above models accessed via Amazon Bedrock. We report the migration efficacy in \cref{tab:os}.


\subsection{Discussion}
\label{subsec:discussion}

\subsubsection{Minimal and Maximal Migration Efficacy.}
{Table \ref{tab:efficay_main} clearly
shows a non-trivial gap between $\eta_\etamin$ and $\eta_\etamax$.}
For Strands agent with \claude\ model,
$\eta_{\etamin}$ is ${4.7}\times$
of
$\eta_{\etamax}$. 
It implies that though minimal migration is relatively straightforward,
maximal migration is much more challenging,
since dependency packages may have to go through major version upgrades and are likely to introduce
\dblue{deprecated or breaking APIs with the same dependency,
or even deprecated or completely renamed dependencies.}

{Comparative analysis based on the Claude model
is shown in \cref{tab:dataset_stats},
demonstrating that
maximal migration 
is more challenging
for larger repositories in terms of number of modules and number of Java files.
}

\subsubsection{Efficiency Study {of $\eta_{\etamax}$}.}
We study the efficiency of the different approaches introduced in \cref{sec:agent} by jointly analyzing maximal migration efficacy and the associated \# LLM {calls}.
Since {the PE agent, RAG agent, and the hybrid approach}
are explicitly designed to solve the maximal migration problem, they are evaluated only under the maximal migration setting {in \cref{tab:efficay_main}}.

{\cref{tab:efficay_main} reports maximal migration efficacy together with the average number of LLM calls per repository.}
The purely static OpenRewrite baseline incurs zero LLM cost but achieves only 2.00\% maximal migration efficacy, indicating its limited ability to handle complex migration issues.
The baseline Strands agent improves {maximal efficacy} to 15.33\% at an average cost of 33.68 LLM calls per repository.
Adding prompt engineering substantially boosts {maximal efficacy} to 45.67\%,
albeit with increased LLM usage.
Incorporating RAG further improves {maximal efficacy} to 53.33\%,
demonstrating the importance of external dependency knowledge for solving advanced migration failures.
The hybrid method achieves the same maximal efficacy (53.33\%) as the RAG agent
{with less LLM calls,}
highlighting {effective cost reduction when} combining deterministic tooling with agentic reasoning.

Figure~\ref{fig:efficacy} illustrates maximal migration efficacy as a function of the number of LLM calls. The baseline Strands agent quickly plateaus at low efficacy, indicating diminishing returns from additional interaction steps. Prompt engineering significantly improves early-stage efficacy, achieving strong gains with relatively fewer calls. RAG and the hybrid approach both exhibit more gradual but sustained improvements, ultimately reaching the highest efficacy levels. Notably, the hybrid method achieves competitive efficacy with fewer agent-driven reasoning steps in early stages, reflecting the benefit of offloading dependency upgrades to static analysis before invoking the agent.

Overall, these results reveal a clear trade-off between migration quality and LLM cost. While richer agentic capabilities lead to higher maximal migration efficacy, the hybrid approach offers a favorable balance by achieving state-of-the-art efficacy with controlled LLM usage, making it a practical and cost-effective solution for large-scale Java migration.

\subsubsection{Compilation Efficacy.}
Recall that we remove all repositories without test cases from \migrationfull, as described in \textbf{\filter{6}} of \cref{subsec:data_collection}. This design choice is motivated by the critical role of unit tests in ensuring functional equivalence before and after migration.

To further support this claim, we conduct an ablation study in which all unit tests are disabled to examine whether migration efficacy increases. Concretely, we replace the default Maven command in $r_1$ with \mvncompile, which only checks whether the source code compiles successfully and does not evaluate test correctness. Under this relaxed criterion, the minimal migration efficacy increases substantially, from \textbf{71.67\%} to \textbf{97.67\%}.

\section{Conclusion}
\label{sec:conclusion}

We introduce \migrationbench,
a comprehensive code migration benchmark dataset from \javan{8} to subsequent LTS versions like \javan{17} and $21$,
with both a \cmd{full} dataset and its \lite\ subset containing $5,102$ and $300$ \repos\ respectively,
\dblue{reflecting real-world code migration scenarios}.
\dblue{We design a comprehensive evaluation framework
to approximate functional equivalence and rigorously assess migration success.
This framework encompasses key criteria,
including \cmd{maven} validations, 
verification of compiled class major versions,
consistency in the list of test methods,
non-decrease in the number of test cases
and adherence to optional dependency version requirements (particularly relevant in maximal migration scenarios).}
We demonstrate the feasibility of code migration from \javan{8} to $17$ through an agentic workflow with
{a few variations of the Strands agents (standalone, + PE, + PE + RAG, and 
a more cost-effective hybrid approach)}, 
and show preliminary results with promising efficacy
for both minimal ($\mineta\%$) and maximal ($\maxeta\%$) 
migration for the \lite\ subset with \claude.
We envision \migrationbench\ as a valuable resource for LLM practitioners and researchers,
fostering further exploration and innovation in coding tasks.
\section{Limitations}
\label{sec:limitations}


We summarize a few limitations of the current work below.
\paragraph{Test Coverage.}
We don't enforce test coverage
and fully reply on \textit{existing} test cases,
although we do require the existence of test cases or dedicated test directory in source language $L_S$.
Functional equivalence has a stronger guarantee when both the source code and tests are migrated to and pass under the target language $L_T$.
However, for lines without any coverage, \maven's \cmd{verify} command is equivalent to \cmd{compile} only.
Fortunately,
our \cmd{UTG} subset can fill in the gap, serving
as a benchmark for unit test generation. We leave unit test generation with \cmd{UTG} subset as a promising future research direction.


\paragraph{Throttling with \cmd{mvn}}
We \textit{completely} rely on running verifiers (e.g.\ various \cmd{maven} commands)
when searching for the base commit ID $H_b$,
running agentic framework and conducting the final evaluation.
However,
we encounter severe throttling
when there are too many concurrent requests to maven central,
therefore it's not rare to introduce false negatives and 
reproducibility issues.
While this is not a major issue with the \lite\ subset,
it remains a challenge when searching for $H_b$ to get the \cmd{full} subset with AWS EMR service, and
we might end up with a base commit other than the actual $H_b$
or completely miss a \repo.

\newpage
{
\bibliographystyle{plain}
\bibliography{reference}
}

\appendix

\section{Data Collection}

We show filter details for \migrationbench\ dataset collection,
to augment \cref{subsec:data_collection}.

\subsection{\filter{4}: Search for Base Commit ID $H_b$}
\label{app:subsec:base_commit_search_appendix}


When we go through the \repo's commit history in the reverse order up to March $2024$,
we assume both \java\ versions and compiled class major versions are chronologically monotonic,
which holds for most \repos.
For initial hard coded java versions, 
while larger than \javan{8},
we fully leverage commit history for \cmd{pom.xml} files \textit{only}
to accelerate identifying the last commit with \javan{8},
instead of going through commit history for the whole \repo.
Whenever we encounter either hard coded \java\ versions or
complied classes' major versions are less than the specification before finding a valid base commit,
we terminate the search and exclude the \repo\ from the benchmark dataset.

When it takes too long to run \mvnverify\ 
or a \repo\ is in a broken state with too many commits,
and we're unable to find a valid base commit within $15$ minutes,
we also exclude it from \migrationbench.
This is the best effort to keep the process to be more reproducible,
take less compute time
and preserve as many \repos\ as possible.

Due to the importance of base commit id $H_b$,
for the \textit{whole} \migrationbench\ dataset including both the \cmd{full} and \cmd{UTG} subsets,
we re-run the process twice and only keep the ones reporting the same $H_b$ across the two runs.
Though we may miss \repos\ due to inconsistency introduced by
e.g.\ \maven\ throttling issues and runtime cutoff,
it provides a \textit{stronger} guarantee on both correctness and reproducibility.

\section{Benchmark Dataset Statistics: \cmd{full} and \lite\ Comparison}
\label{sec:appendix_historgram}

\begin{figure*}[t]
\begin{center}
\includegraphics[width=0.72\textwidth]
{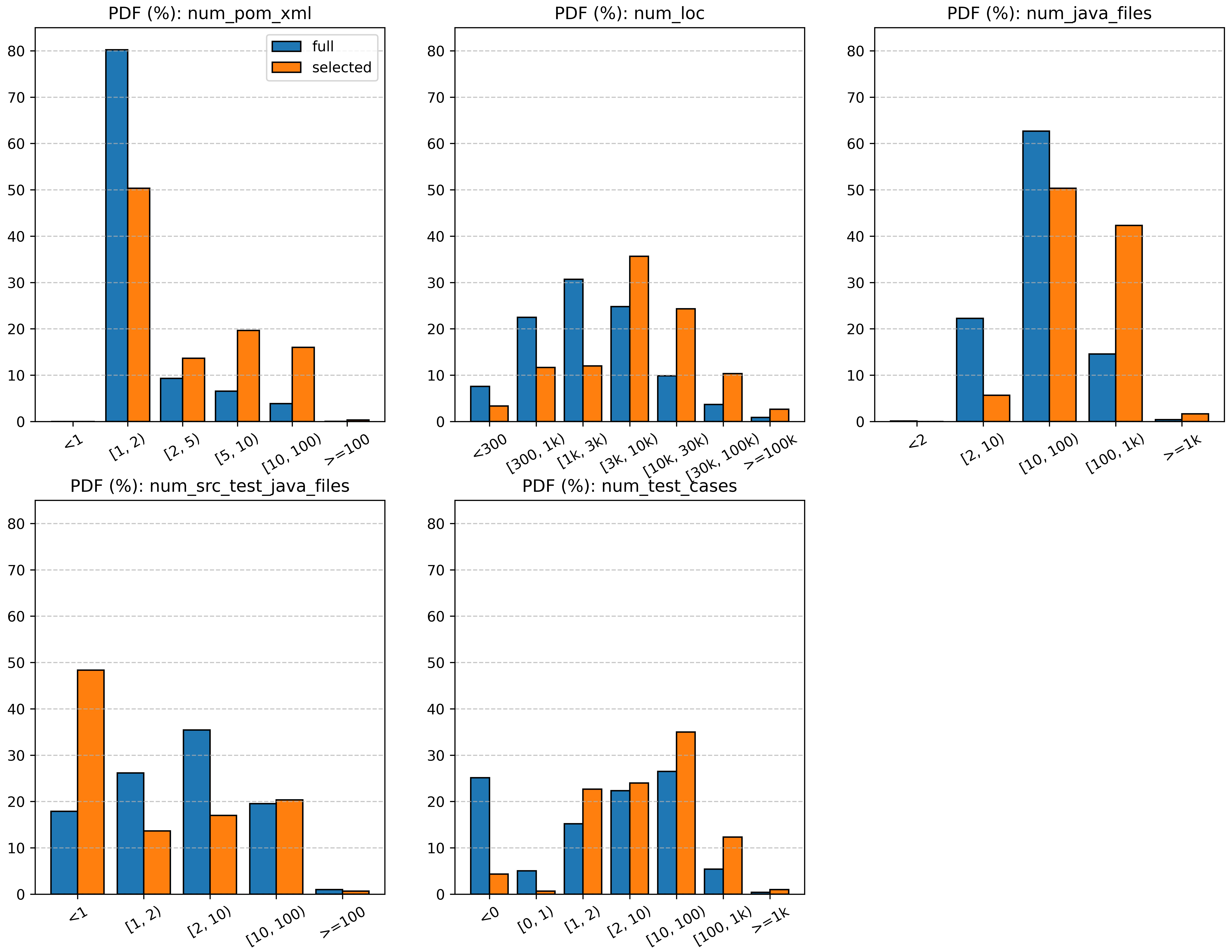}
\end{center}
\caption{Discrete probability density function comparison for
number of modules, LOC, all or test \java\ files and test cases for \repos\ in \migrationbench's \cmd{full} and \lite\ subsets.}
\label{fig:appendix_hist}
\Description{Discrete probability density function comparison for Github repositories statistics in two sets.}
\end{figure*}

\begin{figure*}[t]
\begin{center}
\includegraphics[width=0.72\textwidth]
{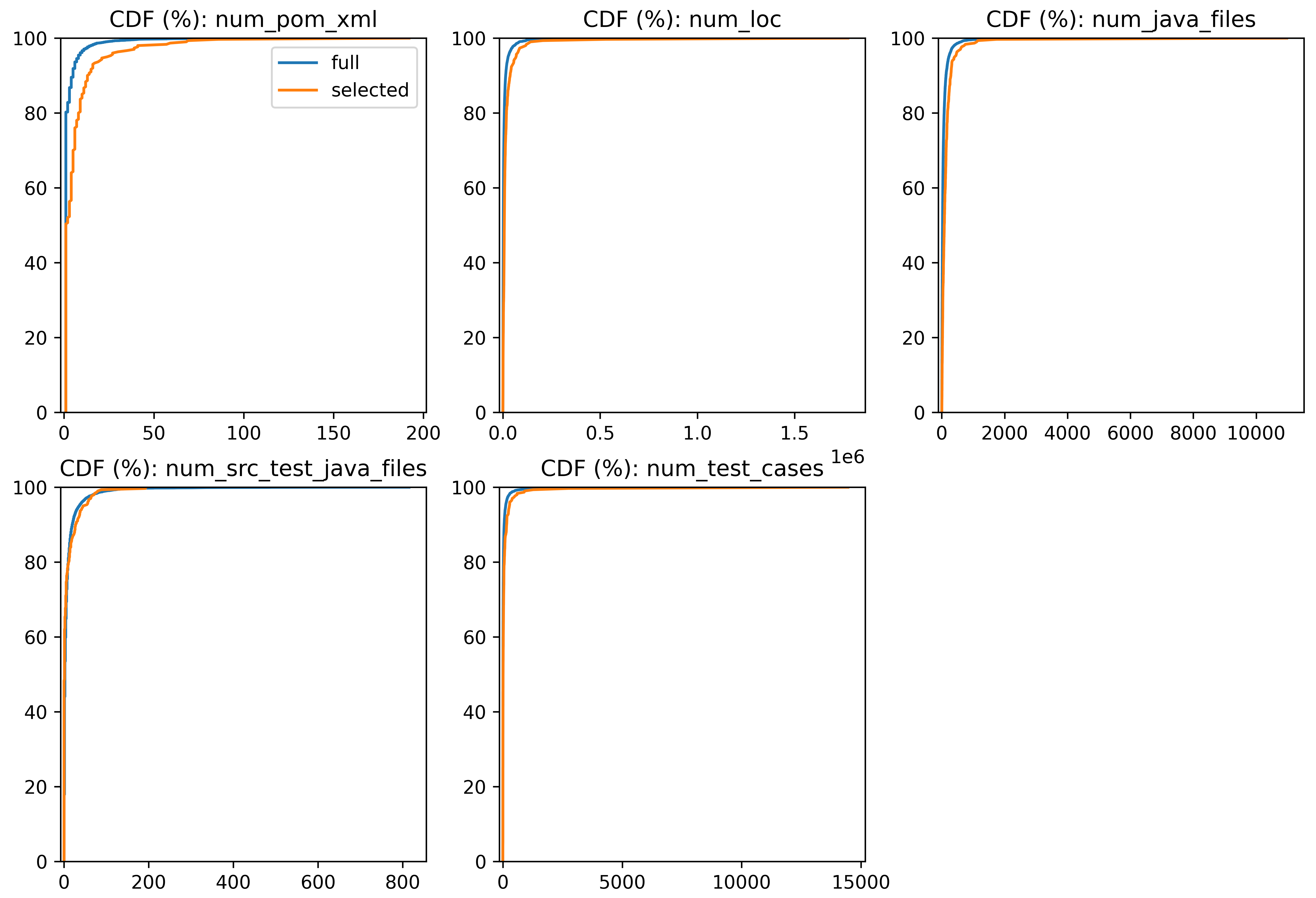}
\end{center}
\caption{Cumulative distribution function comparison for number of modules,
LOC,
all or test-only \java\ files
and test cases \dblue{(dropping invalid negative values)}
for \repos\ in \migrationbench's \cmd{full} and \lite\ subsets.}
\label{fig:appendix_cdf}
\Description{Cumulative distribution function comparison for Github repositories statistics in two sets.}
\end{figure*}

In addition to dataset statistics shown in \cref{subsec:migration-bench-stats}'s \cref{tab:dataset_stats},
in \cref{fig:appendix_hist,fig:appendix_cdf}
we compare the
discrete probability distribution function (PDF) and
cumulative distribution function (CDF) respectively,
for \repos\ in \migrationbench's \cmd{full} and \lite\ subsets:
\begin{enumerate}
    \item Number of modules:
    It shows number of \cmd{pom.xml} files for each \repo.
    \item LOC: {It's a sum of total LOCs}
    for all \cmd{*.java} files for each \repo.
    \item Number of all or test-only \java\ files:
    It counts all or test-only (in the \cmd{src/test} directory)
    \cmd{*.java} files in a \repo.
    \item Number of test cases:
    \dblue{It parses number of test cases based on \cmd{mvn test -f .} command's output at the root directory,
    valid when non-negative,
    while a negative number means there are some issues running the tests
    or parsing the output,
    likely due to \cmd{maven} throttling issues though we try to mitigate it with multiple efforts.}
\end{enumerate}







\section{Evaluation}
\label{sec:appendix_eval}

This section provides more details and extends \cref{sec:eval}.





\subsection{Unit Test Coverage}
\label{subsec:appendix_eval_req_ut}


Unit testing plays a pivotal role in ensuring functional equivalence during code migration,
as it provides a robust mechanism to verify that the migrated code performs identically to its original counterpart,
independent of whether they contain bugs or not.
By isolating individual components of the code and testing them against predefined inputs and expected outputs,
unit tests captures
discrepancies introduced during migration,
such as logic errors,
incompatibilities
or unintended side effects.
This granular validation process builds confidence that the \repo's new state in $L_T$ retains the intended functionality as $L_S$ 
and facilitates early detection and resolution of inconsistency issues.

Passing existing unit tests is already ensured in requirement $r_1$ by executing \mvnverify.
However,
it is strongly recommended to generate additional unit tests to enhance test coverage,
especially when the existing coverage is very low. 

\section{Additional Experiments}
We also perform additional experiments on a sampled subset of \migrationfull\ to highlight that the \cmd{selected} subset is indeed a more challenging one. In this experiment, we sampled every 18 \repos\ from \migrationfull, resulting in 283 \repos. We run hybrid approach with \cmd{Claude-4-5-Sonnet} on this sampled set.

\begin{table}[!ht]
\caption{Maximal migration efficacy on a sampled subset of \migrationfull. We use hybrid approach with \cmd{Claude-4-5-Sonnet}. }
\label{tab:full}
\begin{center}
\scalebox{1}{
\begin{tabular}{lcc}
\toprule
Dataset  & $\eta_{\etamax}$ (\%) $\uparrow$ & Avg. \# LLM Calls per Repo $\downarrow$\\
\midrule
sampled full & 72.79 & 42.30\\
\cmd{selected} & 53.33 & 52.55\\
\bottomrule
\end{tabular}
}
\end{center}
\end{table}

The experiments from \cref{tab:full} verified that the selected subset is more challenging in that it has lower success rate and that it requires more LLM calls to complete.

\end{document}